\documentclass[12pt]{article}
\usepackage{graphicx}

\newcommand\pubnumber{}
\newcommand\pubdate{\today}

\def\hh{Institut f\"ur Experimentalphysik\\
Universit\"at Hamburg, D-22607 Hamburg, Germany}

\def\support{\footnote{Work supported by the German Federal Ministry of Science (BMBF), under project 05H15GUCC1}}

\def\Title#1{\begin{center} {\Large #1 } \end{center}}
\def\Author#1{\begin{center}{ \sc #1} \end{center}}
\def\Address#1{\begin{center}{ \it #1} \end{center}}

\newcommand\pubblock{\rightline{\begin{tabular}{l} \pubnumber\\
         \pubdate  \end{tabular}}}
\newenvironment{Abstract}{\begin{quotation}  }{\end{quotation}}
\newenvironment{Presented}{\begin{quotation} \begin{center} 
             PRESENTED AT\end{center}\bigskip 
      \begin{center}\begin{large}}{\end{large}\end{center} \end{quotation}}
\def\Acknowledgements{\bigskip  \bigskip \begin{center} \begin{large}
             \bf ACKNOWLEDGEMENTS \end{large}\end{center}}

\def\beq{\begin{equation}}
\def\eeq#1{\label{#1}\end{equation}}
\def\eeqn{\end{equation}}

\def\beqa{\begin{eqnarray}}
\def\eeqa#1{\label{#1}\end{eqnarray}}
\def\eeqan{\end{eqnarray}}

\let\bar=\overbar

\def\Dslash{\not{\hbox{\kern-4pt $D$}}}
\def\dslash{\not{\hbox{\kern-2pt $\del$}}}

\def\msb{{\bar{\ssstyle M \kern -1pt S}}}

\begin{document}
\begin{titlepage}
\pubblock

\vfill
\Title{Boosted Top Tagging Method Overview}
\vfill
\Author{Gregor Kasieczka\support}
\Address{\hh}
\vfill
\begin{Abstract}

We briefly review  common tools and methods to identify
boosted, hadronically decaying top quarks at the LHC experiments. This includes
generic jet substructure variables, specific top identification
algorithms, and recent developments in deep learning techniques.

\end{Abstract}
\vfill
\begin{Presented}
TOP 2017\\
Braga, Portugal,  September 17--22, 2017
\end{Presented}
\vfill
\end{titlepage}
\def\thefootnote{\fnsymbol{footnote}}
\setcounter{footnote}{0}

\section{Introduction}

While the discovery of the Higgs boson in 2012 marked the completion
of the Standard Model (SM) as we know it today, important questions such as
the nature of dark matter, the origin of the matter-antimatter
imbalance, the stability of the electroweak vacuum, and the relative
lightness of the Higgs boson remain open.

In answering these questions, many theories beyond the SM
predict new massive particles to be produced at the LHC. Their
subsequent decays may contain, among other signatures, top quarks with
$p_T/m > 1$. If these \textit{boosted} top quarks decay hadronically, their decay products
become collimated and can be experimentally reconstructed as a jet
with large distance parameter (\textit{large-R jet}).

Distinguishing such top decays from jets initiated by light quarks or
gluons is the goal of top tagging algorithms.  In the following we
review three broad classes of such algorithms: generic
QCD inspired jet substructure variables, inclusive top tagging
algorithms, and machine learning approaches\footnote{The reconstruction of b quarks inside a candidate jet is another
handle to identify the decays of top quarks. The current experimental
approach is to largely treat this as an independent problem and we
will not discuss it further. However, the recent advent of
maximum-information machine learning techniques will likely blur this
distinction.}.

While this document can serve as a starting point, the following
reviews as well as proceedings of the annual BOOST conference provide a more
thorough survey of the
field~\cite{BoostReport2016Theory,BoostReport2012,TilmanTagsTops}.

\section{QCD Inspired Jet Substructure Variables}

Over the last decade, a number of techniques 
--- known as \textit{grooming algorithms} ---
have been developed to remove radiation due to the underlying event and pile-up from jets while extracting
information about the hard substructure, especially its mass.

Trimming~\cite{Trimming} first reclusters the constituents of a given jet
with a smaller distance parameter $R_{\textrm{sub}}$ resulting in a
set of so-called subjets. In a second step all subjets with a
transverse momentum of less than a fraction $f_\textrm{cut}$ of the
initial jet transverse momentum are discarded. Finally, the trimmed
jet is obtained by summing up the remaining subjets. Trimming 
is the default
grooming algorithm employed by the ATLAS collaboration~\cite{CONF-17-063}.

Conversely, the soft drop algorithm~\cite{SoftDrop} attempts to find the hard 
substructure using a step-wise unclustering procedure. The initial jet is recursively unclustered until 
no constituents remain or until
the soft drop condition
\begin{equation}
\frac{\min(p_{T1},p_{T2})}{p_{T1}+p_{T2}} > z_{cut} \left( \frac{\Delta R_{12}}{R_0}\right)^\beta
\end{equation}
is fulfilled. 
Here $p_{T1}$ ($p_{T2}$) denotes the transverse momentum of the subjet with higher (lower) transverse momentum, 
$\Delta R_{12}$ their angular distance and $z_{cut}$, $R_0$, and $\beta$ 
are free parameters of the algorithm. 
The recursion proceeds at each step by discarding the lower momentum subjet and unclustering the higher momentum one. For $\beta=0$ the algorithm corresponds to the modified mass drop algorithm~\cite{MMDT}. The soft drop algorithm is widely used by the CMS collaboration~\cite{JME-16-003}. Recent attempts have been made to analytically calculate the properties of soft dropped jets to high precision~\cite{SoftDropFact,SoftDropMassCalcI,SoftDropMassCalcII} and to perform corresponding measurements with LHC collision data~\cite{PAS-SMP-16-010,ATLASSDMeas}.

Beyond the mass, the number of distinct centers of radiation is well
suited to identify top quarks (with three centers) from other objets
(which are expected to have one or two). One way to quantify this intuition is given by the n-subjettiness~\cite{NSubjettiness} $\tau_N$.

The n-subjettiness is small
when the radiation inside a jet is compatible with $N$ hard centers of
radiation. Commonly the ratio $\tau_3/\tau_2$ is used for top
tagging. Further developments include the dichroic n-subjettiness~\cite{DichroicNSubjettiness} ---
the ratio of n-subjettiness values after different grooming techniques
are applied --- and its use for jet clustering~\cite{XCone}.

More recently, energy correlation functions have been proposed which
generalize the n-subjettiness by calculating n-particle correlators
and therefor do not rely on the explicit finding of subjet axes~\cite{ECF,ECF2}.

\section{Inclusive Top Tagging Algorithms}

In addition to the generic variables discussed in the previous
section, a number of inclusive top tagging algorithms have been
developed.

The HEPTopTagger~V2~\cite{HTT,HTTv2} algorithm identifies top quarks by a mass-drop
filtering procedure followed by selection criteria on the top mass and
W to top mass ratio. This procedure is repeated for a number jet of
distance parameters, starting from a seed size of $R=1.5$ until the
optimal jet size is found. 

The Heavy Object Tagger with Variable R~\cite{HOTVR} (HOTVR) also implements jet
clustering with variable distance parameter. Soft radiation is already removed during
the clustering procedure, using a mass drop criterion leading to a
reconstructed jet mass stable over a large transverse momentum range.

Shower deconstruction~\cite{ShowersI,ShowersII} is a full information approach similar to the
matrix element method that assigns signal and background
probabilities to individual jets instead of events. First, microjets
are built by clustering the constituents of the large-R jet into
smaller jets.
The classifying variable
$\chi$ is then calculated as the probability quotient that a set of
microjets was created by the decay of a top quark, divided by the
probability that it was created by light quarks or gluons.

The Template Overlap Method~\cite{TemplateI,TemplateII} uses a library of templates which encode
the distribution of parton momenta inside a jet. The classifier is
then calculated as the maximum agreement between the jet under study
and all templates for a given phase space.

Finally, the Boosted Event Shape Tagger~\cite{BEST} (BEST) starts by boosting jet
constituents individually into reference frames corresponding to a
particle origin hypothesis (t, W, Z, H). Subsequently, angular
distributions such as 
Fox-Wolfram moments or sphericity are calculated in each
frame.  A fully connected neural network then yields the compatibility
with each particle hypothesis for simultaneous classification.

\section{(Deep) Machine Learning Approaches}

Deep learning is a novel development in computer science, largely
based on creating artificial neural networks that feature multiple
layers and correspondingly increasingly higher levels of
abstraction. Neural networks connect a number of inputs to an output
decision via intermediate nodes. The output of a given node is
determined by its inputs and a set of weights that are adjusted
(trained) for a given decision task. How many internal nodes to use
and how to connect them is referred to as the
network architecture. In addition to using large numbers of internal
layers, deep network approaches typically operate directly on
minimally preprocessed input data (as opposed to variables designed by
experts discussed in the previous sections).

An important decision when designing a deep neural network for a specific
task is the choice of the architecture. In the following we outline 
several network architectures used for top tagging.

In a fully connected network (FCN), each node in a given layer is connected
to all nodes of the previous and subsequent layers. This architecture
is very generic, but does not make explicit use of symmetries inherent
to the problem at hand. An application of FCNs to
top tagging is presented in~\cite{FCN}. The individual constituents of a jet
are given to a FCN with four hidden layers. Several steps of pre-processing
including trimming and scaling, translation, rotation, and flipping
steps increase the achieveable performance.

Image recognition tasks are commonly solved using 
convolutional neural networks (CNNs). In these networks the
weights to be learned are arranged in a so-called convolutional kernel
(or several thereof) that is applied to different parts of the
image. This approach can thus detect features useful for
discrimination no matter where on the image they occur. 
Top tagging can be viewed as an image
recognition task by equating the energy deposits in the
calorimeters with the pixels of a grayscale image~\cite{DeepTop}.

An alternative approach, inspired by the processing of human language
are recursive neural networks (RNNs). A sentence (sequence)
consists of an arbitrary number of words (inputs). RNNs process the
elements of an input sequence one after the other. Each input in the
sequence is handled by the same processing layer (thus
recursive). Importantly, RNNs also possess internal memory. 
Recently, a combination of RNNs and jet
clustering algorithms was used to develop a competitive identification
technique for boosted top quarks~\cite{LSTM}.

Finally, the physics inspired LoLa algorithm~\cite{LoLa} starts with a list of input
four vectors which can either be calorimeter towers, particle flow
candidates or generated partons. In a first step a combination layer
(CoLa) yields a set of trainable linear combinations of
the inputs. In a second step, the eponymous Lorentz layer (LoLa) uses
the resulting matrix to extract physics features based on convolutions
with the Minkowski metric. Its output is then passed to a FCN to identify top quarks.

\section{Conclusions}

A varied menu of top tagging techniques is available for use in
physics analyses. While developments previously focused on searches for new
physics, recent precision calculations might make substructure
increasingly interesting for SM measurements. At the same time, deep learning based tagging algorithms 
already show promising gains in performance.


\Acknowledgements
The author would like to thank the organisers of TOP 2017 for the invitation to this wonderfully organised and stimulating conference.


\begin{thebibliography}{99}




\bibitem{BoostReport2016Theory}
  A.~J.~Larkoski, I.~Moult and B.~Nachman,
  arXiv:1709.04464 [hep-ph].

\bibitem{BoostReport2012}
  A.~Altheimer {\it et al., BOOST2012},
  Eur.\ Phys.\ J.\ C {\bf 74} (2014),  2792.

%


\bibitem{TilmanTagsTops}
  T.~Plehn and M.~Spannowsky,
  J.\ Phys.\ G {\bf 39} (2012) 083001.

\bibitem{Trimming}
  D.~Krohn, J.~Thaler and L.~T.~Wang,
  JHEP {\bf 1002} (2010) 084.


\bibitem{CONF-17-063}
  ATLAS collaboration,
  ATLAS-CONF-2017-063.


\bibitem{SoftDrop}
  A.~J.~Larkoski, S.~Marzani, G.~Soyez and J.~Thaler,
  JHEP {\bf 1405} (2014) 146.


\bibitem{MMDT}
  M.~Dasgupta, A.~Fregoso, S.~Marzani and G.~P.~Salam,
  JHEP {\bf 1309} (2013) 029.


\bibitem{JME-16-003}
  CMS Collaboration,
  CMS-PAS-JME-16-003.


\bibitem{SoftDropFact}
  C.~Frye, A.~J.~Larkoski, M.~D.~Schwartz and K.~Yan,
  JHEP {\bf 1607} (2016) 064.

\bibitem{SoftDropMassCalcI}
  S.~Marzani, L.~Schunk and G.~Soyez,
  JHEP {\bf 1707} (2017) 132.

\bibitem{SoftDropMassCalcII}
  S.~Marzani, L.~Schunk and G.~Soyez,
  arXiv:1712.05105 [hep-ph].


\bibitem{PAS-SMP-16-010}
  CMS Collaboration,
  CMS-PAS-SMP-16-010.

\bibitem{ATLASSDMeas}
  ATLAS Collaboration,
  arXiv:1711.08341 [hep-ex].


\bibitem{NSubjettiness}
  J.~Thaler and K.~Van Tilburg,
  JHEP {\bf 1103} (2011) 015.

\bibitem{DichroicNSubjettiness}
  G.~P.~Salam, L.~Schunk and G.~Soyez,
  JHEP {\bf 1703} (2017) 022.

\bibitem{XCone}
  J.~Thaler and T.~F.~Wilkason,
  JHEP {\bf 1512} (2015) 051.

\bibitem{ECF}
  A.~J.~Larkoski, G.~P.~Salam and J.~Thaler,
  JHEP {\bf 1306} (2013) 108.

\bibitem{ECF2}
  I.~Moult, L.~Necib and J.~Thaler,
  JHEP {\bf 1612} (2016) 153.


\bibitem{HTT}
  T.~Plehn, M.~Spannowsky, M.~Takeuchi and D.~Zerwas,
  JHEP {\bf 1010} (2010) 078.


\bibitem{HTTv2}
  G.~Kasieczka, T.~Plehn, T.~Schell, T.~Strebler and G.~P.~Salam,
  JHEP {\bf 1506} (2015) 203.

\bibitem{HOTVR}
  T.~Lapsien, R.~Kogler and J.~Haller,
  Eur.\ Phys.\ J.\ C {\bf 76} (2016),  600.

\bibitem{ShowersI}
  D.~E.~Soper and M.~Spannowsky,
  Phys.\ Rev.\ D {\bf 87} (2013) 054012.

\bibitem{ShowersII}
  D.~E.~Soper and M.~Spannowsky,
  Phys.\ Rev.\ D {\bf 89} (2014),  094005.

\bibitem{TemplateI}
  L.~G.~Almeida, S.~J.~Lee, G.~Perez, G.~Sterman and I.~Sung,
  Phys.\ Rev.\ D {\bf 82} (2010) 054034.

\bibitem{TemplateII}
  M.~Backovic, J.~Juknevich and G.~Perez,
  [arXiv:1212.2977 [hep-ph]].

\bibitem{BEST}
  J.~S.~Conway, R.~Bhaskar, R.~D.~Erbacher and J.~Pilot,
  Phys.\ Rev.\ D {\bf 94} (2016),  094027.


\bibitem{FCN}
  J.~Pearkes, W.~Fedorko, A.~Lister and C.~Gay,
  arXiv:1704.02124 [hep-ex].

\bibitem{DeepTop}
  G.~Kasieczka, T.~Plehn, M.~Russell and T.~Schell,
  JHEP {\bf 1705} (2017) 006.


\bibitem{LSTM}
  S.~Egan, W.~Fedorko, A.~Lister, J.~Pearkes and C.~Gay,
  arXiv:1711.09059 [hep-ex].


\bibitem{LoLa}
  A.~Butter, G.~Kasieczka, T.~Plehn and M.~Russell,
  arXiv:1707.08966 [hep-ph].


\end{thebibliography}
\end{document}